\documentclass[preprint,english,showpacs,11pt,floatfix]{revtex4}
\def\simgreat{\mathbin{\lower 3pt\hbox
   {$\rlap{\raise 5pt\hbox{$\char'076$}}\mathchar"7218$}}}

\def\b2{\beta_2}

\def\a2{\alpha_2}

\def\be{\begin{equation}}
\def\ee{\end{equation}}
\def\baray{\begin{eqnarray}}
\def\earay{\end{eqnarray}}

\usepackage{babel,amsmath,amssymb,dcolumn}
\usepackage[dvips]{graphics}
\usepackage{hyperref}

\usepackage{amsfonts}
\usepackage{amssymb}
\usepackage[dvips]{graphicx}
\usepackage{latexsym}

\usepackage{dcolumn}
\usepackage{bm}

\begin{document}

\title{Super-acceleration on the Brane by Energy Flow from the Bulk }
\author{Rong-Gen Cai\footnote{Email address: cairg@itp.ac.cn}$^{,1}$, Yungui Gong\footnote{
Email address: gongyg@cqupt.edu.cn}$^{,2,3}$
and Bin Wang\footnote{Email address: wangb@fudan.edu.cn}$^{,4}$
}
\affiliation{
$^1$Institute of Theoretical Physics, Chinese Academy of Sciences, \\
P.O. Box 2735, Beijing 100080, China\\
$^2$College of Electronic
Engineering,\\
Chongqing University of Posts and Telecommunications, \\
Chongqing 400065, China \\
$^3$CASPER, Physics Department, Baylor University, Waco, TX 76798, USA\\
$^4$Department of Physics,
Fudan University, Shanghai 200433,  China \\
}

\begin{abstract}
We consider a brane cosmological model with energy exchange
between brane and bulk. Parameterizing the energy exchange term by
the scale factor and Hubble parameter, we are able to exactly
solve the modified Friedmann equation on the brane. In this model,
the equation of state for the effective dark energy has a
transition behavior changing from $w_{de}^{eff}>-1$ to
$w_{de}^{eff}<-1$, while the equation of state for the dark energy
on the brane has $w>-1$. Fitting data from type Ia supernova,
Sloan Digital Sky Survey and Wilkinson Microwave Anisotropy Probe,
our universe is predicted now in the state of super-acceleration
with $w_{de0}^{eff}=-1.21$.

\end{abstract}


\maketitle

A variety of cosmological observations consistently provide a
compelling evidence that our universe is experiencing an
accelerated expansion. A component that causes the expansion of
the universe to accelerate is referred as dark energy. The
traditional cosmological constant is a possible candidate of the
dark energy, although the 120 orders of magnitude difference
between its theoretical and observational values presents the
biggest problem to theorists~\cite{01}. Another popular candidate
is an exotic field with evolving equation of state~\cite{02}.
However compared to the comfortable vacuum energy interpretation
of the cosmological constant, the natural reason behind these
exotic fields is lacking. There are also alternative models of
gravity that seek to explain the accelerated
expansion\cite{03,04,05,06,07,08,09,10,in1,in2}. One interesting
theory is the so-called $1/R$ gravity suggested in \cite{03,09},
which is possible to account for the late time acceleration. While
unfortunately this theory was argued suffering a conflict with
gravitational tests in the solar system \cite{11}. Recently an
alternative way to avoid this conflict was proposed \cite{12}.
Another dramatic strategy to modify gravity is to imagine that we
live on a brane embedded in a higher dimensional spacetime
\cite{05,10,in2,RSII}, which can naturally lead to late time
acceleration. In these studies, it was assumed that dark energy
does not interact with matter or radiation.

Given the unknown nature of both dark energy and dark matter,
which are two major contents of the universe, one might argue that
an entirely independent behavior of dark energy is very special
\cite{13,14}. A lot of investigations assuming interaction between
dark energy and dark matter have been carried out \cite{15,16}.
With the interaction between dark sectors, the standard
conservation equations of matter and dark energy
 are violated respectively. While the Friedmann equations can be
easily modified to model exchanges between different energy
components, studies on dark energy models with interaction with
matter fields have disclosed that the equation of state of the
dark energy has a transition behavior, changing from $w_D>-1$ to
$w_D<-1$ at recent stage \cite{15,16}. This theoretical result is
consistent with the recent extensive analysis of observation data
\cite{17}. The energy exchanges between dark sectors will drive
the universe to super-acceleration \cite{15,16,18} and in addition
it will influence the perturbation dynamics and could also be
observable through the lowest multipoles of Cosmic Microwave
Background (CMB) spectrum \cite{14}. Besides the energy exchange
between dark sectors considered in four-dimensional theories, the
role of the energy exchange between the bulk and brane has also
been investigated. The cosmic evolution of the brane in the
presence of energy flow into or from the bulk has been analyzed in
different setups \cite{19,20}.

The dark energy with equation of state $w<-1$ is referred as
phantom dark energy~\cite{Cald}. An easy way to realize the
phantom dark energy is a scalar field with a wrong sign kinetic
term. However, such a model might suffer from the
instability~\cite{Carr}. A more challenging issue is that the time
dependent dark energy gives a better fitting than a cosmological
constant, and in particular, the equation of state crosses $-1$ at
redshift $z\approx 0.2$ from the above to below~\cite{in3}. While
it turns out it is not trivial to build dark energy model with
equation of state crossing $w=-1$~\cite{Vik}, a component with
$w<-1$ in the universe will violate all energy conditions and
threaten the basis of modern physics. Therefore it is urgent and
is of great interest to construct dark energy model with observed
feature in the framework of modern physics. That is, one may hope
that crossing $-1$ for the equation of state and phantom behavior
of dark energy are effective (or equivalent) features, not caused
by an unstable and causality-violated phantom field. Indeed, it
can be done in the scalar-tensor theory~\cite{in1}, brane world
scenario~\cite{Cai1}, and models with interactions between dark
matter and dark energy, and so on.

In this paper, we will consider a brane world scenario (RSII
model)~\cite{RSII}, in which to realize the transition of $w$ from
above $-1$ to below $-1$ by introducing an energy exchange between
the brane and the bulk. By approximately parameterizing the energy
exchange, we are able to exactly solve the resulting Friedmann
equation on the brane. The ``phantom" behavior of the dark energy
on the brane is caused by energy flow from the bulk. Both dark
energies on the brane and from the bulk obey the causality
condition.  In this sense the transition of the effective dark
energy equation of state could serve as a signature of the
bulk-brane energy exchange.

The gravitational brane-bulk action we are going to consider is of
the form
\begin{equation}\label{eq1}
    S=\int d^5 x
    \sqrt{-G}\left(\frac{R_5}{2\kappa_5^2}-\Lambda_5+\mathcal{L}_B^m \right)+ \int
    d^4 x \sqrt{-g}\left(-\sigma + \mathcal{L}_b^m \right),
\end{equation}
where $R_5$ is the curvature scalar of the five-dimensional
metric, $\Lambda_5$ is the bulk cosmological constant and $\sigma$
is the brane tension, $\mathcal{L}_B^m$ and $\mathcal{L}_b^m$ are the
matter Lagrangian in the bulk and on the brane respectively. We are interested in the cosmological
solutions with a metric
\begin{equation}\label{eq2}
    ds^2 = -n^2(t,y)\,dt^2 + a^2(t,y)\,\gamma_{ij}\,dx^i dy^j +
    b^2(t,y)\,dy^2.
\end{equation}
The non-zero components of Einstein tensor can be written as \cite{21}
\baray && G_{00}=3 \left[ \frac{ \dot a}{a} \left( \frac{ \dot
a}{a} + \frac{ \dot b}b \right) - \frac{n^2}{b^2} \left( \frac{a^{
\prime \prime }}a + \frac{a^{ \prime }}a \left( \frac{a^{
\prime}}a- \frac{b^{\prime}}b \right)
\right) + k \frac{n^2}{b^2} \right], \\
&& G_{ij}= \frac{a^2}{b^2}\gamma_{ij} \left[\frac{a^{ \prime}}a
\left( \frac{a^{ \prime}}a+2\frac{n^{
\prime}}n\right)-\frac{b^{\prime}}b\left(
\frac{n^{\prime}}n+2\frac{a^{ \prime}}a\right)+2\frac{a^{\prime
\prime}}a+ \frac{n^{\prime \prime}}n\right] \nonumber \\
&& ~~~~~+\frac{a^2}{n^2}\gamma_{ij}\left[ \frac{\dot
a}a\left(-\frac{\dot a}a+2\frac{\dot n}n\right)-2\frac{\ddot a}a
+\frac{\dot b}b\left(-2\frac{\dot a}a+\frac{\dot n}n
\right)-\frac{\ddot b}b
\right] -k\gamma_{ij}, \\
&& G_{05}=3\left( \frac{n^{\prime}}n\frac{\dot a}a
+\frac{a^{\prime}}a\frac
{\dot b}b - \frac{\dot a^{\prime}}a \right), \\
&& G_{55}=3\left[\frac{a^{\prime}}a\left(\frac{a^{\prime}}a+
\frac{n^{\prime}}n \right)-\frac{b^2}{n^2}\left(\frac{\dot
a}a\left(\frac{ \dot a}a-\frac{\dot n}n \right)+\frac{\ddot
a}a\right)-k\frac{b^2}{a^2}\right],
 \earay
  where $\gamma_{ij}$ is
the metric for the  maximally symmetric three-dimensional space
and $k = -1, 0, 1$ representing its curvature. In the above
equations, primes and dots stand for derivatives with respect to
$y$ and $t$ respectively. The three dimensional brane is assumed
at $y=0$. The Einstein equations are $G_{\mu \nu}= \kappa^2_5 T_{\mu\nu}$, where
 the stress-energy momentum tensor has  bulk and brane
components and can be written as
\begin{equation}
\label{eq7}
    T^{\mu}_{\phantom{\mu}\nu} = T^{\mu}_{\phantom{\mu}\nu}|_{\sigma,\,b}
    + T^{\mu}_{\phantom{\mu}\nu}|_{m,b} +
    T^{\mu}_{\phantom{\mu}\nu}|_{\Lambda,\,B} +
    T^{\mu}_{\phantom{\mu}\nu}|_{m,B},
\end{equation}
where
\begin{eqnarray}
  T^{\mu}_{\phantom{\mu}\nu}|_{\sigma,\,b} &=& \frac{\delta(y)}{b}\,
  \mathrm{diag}(-\sigma,\,-\sigma,\,-\sigma,\,-\sigma,\,0), \label{eq8}\\
  T^{\mu}_{\phantom{\mu}\nu}|_{\Lambda,\,B} &=& \mathrm{diag}(-\Lambda_5,
  \,-\Lambda_5,\,-\Lambda_5,\,-\Lambda_5,\,-\Lambda_5), \label{eq9}\\
  T^{\mu}_{\phantom{\mu}\nu}|_{m,\,b} &=&
  \frac{\delta(y)}{b}\,\mathrm{diag}(-\rho,\,p,\,p,\,p,\,0),\label{eq10}
\end{eqnarray}
$\rho$ and $p$ are energy density and pressure on the brane,
respectively.

Assuming the $Z_2$ symmetry around the brane, we can obtain
\begin{eqnarray}
  a'_{+} &=& -\,a'_{-} = -\frac{\kappa_5^2}{6}\,a_0\,b_0\,(\sigma + \rho), \label{eq11}\\
  n'_{+} &=& -\,n'_{-} = \frac{\kappa_5^2}{6} \,b_0\,n_0\,(-\sigma + 2\rho
  +3p),
   \label{eq12}
 \end{eqnarray}
by integrating Eqs.~(3) and (4) with respect to $y$ around $y=0$, where the
subscripts ``$+$'' and ``$-$'' stand for $y>0$ and $y<0$
respectively, which represent two sides of the brane. In addition,
as usual, the subscript ``$0$" denotes quantities are evaluated at
$y=0$.

From (5) and (6), we obtain
\begin{eqnarray}
  \frac{n'_0\,\dot{a}_0}{n_0\,a_0} +
    \frac{a'_0\,\dot{b}_0}{a_0\,b_0}-\frac{\dot{a}'_0}{a_0} &=&
    \frac{\kappa_5^2}{3}\,T_{05}, \label{eq13} \\
  3\left\{ \frac{a'_0}{a_0} \left( \frac{a'_0}{a_0} + \frac{n'_0}{n_0} \right)
  - \frac{b^2_0}{n^2_0} \left[ \frac{\dot{a}_0}{a_0} \left( \frac{\dot{a}_0}{a_0}
   -\frac{\dot{n}_0}{n_0} \right)+ \frac{\ddot{a}_0}{a_0}\right]
    -k \frac{b_0^2}{a_0^2} \right\} &=& -\kappa_5^2\,\Lambda_5\,b_0^2 + \kappa_5^2\,T_{55},
     \label{eq14}
\end{eqnarray}
where $T_{05}$ and $T_{55}$ are the 05 and 55 components of
$T_{\mu\nu}|_{m,\,B}$ evaluated on the brane. Employing
Eqs.~(\ref{eq11}) and (\ref{eq12}), we can derive
\begin{eqnarray}
  \dot{\rho} + 3 \frac{\dot{a}_0}{a_0}(\rho+p) &=& -\frac{2
n^2_0}{b_0}T^0_{\phantom{0}5},
  \label{eq15} \\
  \frac{1}{n_0^2} \left[ \frac{\ddot{a}_0}{a_0} + \left( \frac{\dot{a}_0}{a_0} \right)^2
   -\frac{\dot{a}_0\,\dot{n}_0}{a_0\,n_0} \right] + \frac{k}{a_0^2} &=&
  \frac{\kappa_5^2}{3} \left( \Lambda_5 + \frac{\kappa_5^2\,\sigma^2}{6}\right)
  \nonumber\\
    & & - \frac{\kappa_5^4}{36} \left[ \sigma(3p-\rho)+\rho(3p+\rho)
  \right] - \frac{\kappa_5^2}{3}\,T^5_{\phantom{5}5}.
  \label{eq16}
\end{eqnarray}
Taking an appropriate gauge with the coordinate frame $b_0=n_0=1$,
Eqs.~(\ref{eq15}) and (\ref{eq16}) can be reexpressed as
\begin{eqnarray}
  \dot{\rho} + 3 H(1+w)\rho &=& -2\,T^0_{\phantom{0}5}, \label{eq17} \\
  \left( \frac{\dot{a}}{a} \right)^2 &=& \lambda - \frac{\kappa}{a^2} +\beta\,
  \rho^2 + 2\,\gamma\,(\rho + \chi), \label{eq18}\\
  \dot{\chi} + 4H\chi &=& 2\left(\frac{\rho}{\sigma}+1 \right)
  \,T^0_{\phantom{0}5} - \frac{12}{\kappa_5^2} \frac{H}{\sigma}
  \,T^5_{\phantom{5}5}, \label{eq19}
\end{eqnarray}
where $\beta=\frac{\kappa_5^4}{36}$ and
$\gamma=\frac{\sigma\,\kappa_5^4}{36}$. $\lambda \frac{\kappa_5^2}{6} \left( \Lambda_5 +
\frac{\kappa^2_5\,\sigma^2}{6} \right)$ is the effective
cosmological constant on the brane.

In order to derive a solution that is largely independent of the
bulk dynamics, we can neglect $T^5_{\phantom{5}5}$ term by
assuming that the bulk matter relative to the bulk vacuum energy
is much less than the ratio of the brane matter to the brane
vacuum energy~\cite{19}. Considering this approximation and
concentrating on the low-energy region with $\rho / \sigma \ll 1$,
Eqs.~(\ref{eq17})-(\ref{eq19}) can be simplified into
\begin{eqnarray}
  \dot{\rho} + 3 H(1+w)\rho &=& -2\,T^0_{\phantom{0}5} = T \label{eq20} \\
  H^2 &=& \frac{8\pi G_4}{3} (\rho + \chi) - \frac{k}{a^2} + \lambda \label{eq21}\\
  \dot{\chi} + 4H\chi & \approx & 2
  \,T^0_{\phantom{0}5} =-T . \label{eq22}
\end{eqnarray}
Thus with the energy exchange $T$ between the bulk and brane, the
usual energy conservation is broken down. In the following we will
consider that there are two dark components in the universe, dark
matter and dark energy, $\rho = \rho_m + \rho_{de}$. The
bulk-brane energy exchange will break the adiabatic equation
either for the dark matter or the dark energy. We will study these
two cases respectively. In our discussion, we will take
$\lambda=0$ and $k =0$.

\begin{figure}[!hbtp] \label{fig1}
\begin{center}
\includegraphics[width=10cm]{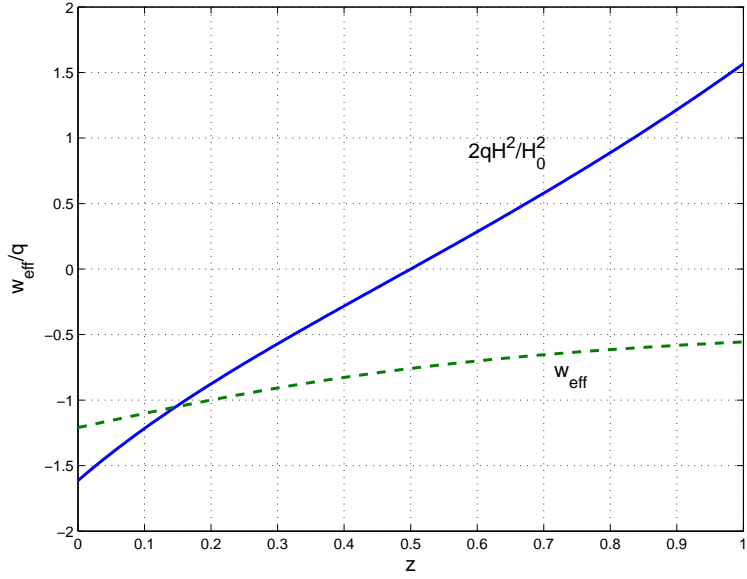}
\end{center}
\caption{Evolution of the deceleration parameter and the equation
of state for the effective dark energy.}
\end{figure}

For the first case, we assume that the adiabatic equation for the
dark matter is satisfied while it is violated for the dark energy
due to the energy exchange between the brane and the bulk,
\begin{eqnarray}
  \dot{\rho}_m + 3 H \rho_m &=& 0, \label{eq23}\\
  \dot{\rho}_{de} + 3 H (1+w) \rho_{de} &=& T. \label{eq24}
\end{eqnarray}
Thus the evolution of dark sectors have the form $\rho_m
=\rho_{m0}/a^3$, and $\rho_{de} = \exp\,[-\int 3 H (1+w)dt]
\left\{ \int T \exp \left[\int 3H(1+w)dt  \right] dt + C
\right\}$. Assuming $w$ being a constant and taking the ansatz
$T=T_0 H a^n$ with $T_0$ and $a$ two constants, we get
$\rho_{de}\frac{C}{a^{3(1+w)}}+ \frac{T_0\,a^n}{n+3(1+w)}$. Some
remarks here are in order. To get a cosmological model based on
Eqs.~(\ref{eq20})-(\ref{eq22}), one has to know the energy
exchange $T$. Unfortunately, it is not yet available and obviously
it depends on  mechanism which produces the energy exchange. Some
references in \cite{19} take the ansatz $T\sim \rho^n$. In that
case, Eqs.~(\ref{eq20}) and (\ref{eq22}) cannot be integrated
analytically and a numerical approach has to be adopted. Some
papers in \cite{19} consider the case without dark energy on the
brane and the late-time acceleration is caused by the effective
cosmological constant, dark radiation and the energy exchange with
the ansatz $T\sim \rho^n$.  Umezu {\it et al.} in \cite {19}
considered the case with $T\sim a^{-n}H^3$. The authors of
\cite{20} generalized these discussions to the case with an
induced gravity term on the brane and with the ansatz $T\sim
\rho^n$. Clearly in those cases, a late-time stable attractor with
accelerated expansion exists. In our setup, both the dark matter
and dark energy with $w>-1$ appear on the brane. The ansatz $T\sim
H a^n$ is taken so that one can directly integrate (\ref{eq20})
and (\ref{eq22}) and obtain their analytic expressions. In
addition, we will consider two cases: one is to transfer the bulk
energy $T$ to the dark energy on the brane; the other is to dark
matter on the brane. These two cases have significant differences
on the cosmic observations.

Now we substitute our ansatz of the bulk-brane energy flow into
Eq.~(\ref{eq22}), and have
\begin{equation}\label{eq25}
    \chi = \frac{C_1}{a^4} - \frac{T_0}{4+n}a^n,
\end{equation}
where $C_1$ is an integration constant. Inserting it into
Eq.~(\ref{eq21}), the Friedmann equation reads
\begin{equation}\label{eq26}
    H^2 = \frac{8\pi G_4}{3} \left[ \frac{\rho_{m0}}{a^3} + \frac{C}{a^{3(1+w)}}
    + \frac{T_0\,(1-3w)a^n}{(4+n)[n+3(1+w)]}\right],
\end{equation}
where we have neglected the dark radiation term $\sim a^{-4}$,
namely $C_1=0$,  since we have more interest in the late time era
of the universe.

Another possibility is to consider transferring the bulk-brane
energy exchange to the dark matter on the brane, and keeping the
standard conservation equation for the dark energy on the brane,
\begin{eqnarray}
  \dot{\rho}_m +3 H \rho_m &=& T, \label{eq27}\\
  \dot{\rho}_{de} + 3 H (1+w) \rho_{de} &=& 0. \label{eq28}
\end{eqnarray}
By assuming $w$ as a constant, and using the ansatz for $T$, the
evolution of dark sectors become $\rho_{de}
=\frac{\rho_{de0}}{a^{3(1+w)}}$ and $\rho_m = \frac{D}{a^3} +
\frac{T_0}{3+n}\,a^n$. From Eq.~(\ref{eq22}), the generalized dark
radiation density can be obtained as $\chi = \frac{D_1}{a^4} -
\frac{T_0}{4+n}\,a^n$. Thus the Friedmann equation has the form
\begin{equation}\label{eq29}
    H^2 = \frac{8\pi G_4}{3}\left( \frac{D}{a^3} + \frac{\rho_{de0}}{a^{3(1+w)}}
    + \frac{T_0\,a^n}{(3+n)(4+n)}
    \right),
\end{equation}
where a term related to the dark radiation has also been neglected
as treated above. That is, $D_1=0$ has been taken.

Thus we have derived the cosmic evolution of the brane in the
presence of energy flow into or from the bulk. By assuming a
specific form of the energy transfer which causes the violation of
the standard conservation equation of either dark matter or dark
energy, we have obtained the modified Friedmann equations
(\ref{eq26}) and (\ref{eq29}). Compared to the usual
four-dimensional theory, we have got additional contributions from
the bulk-brane energy exchanges which are similar in the third
terms of (\ref{eq26}) and (\ref{eq29}). When $w=0$, we notice that
these two cases coincide with each other. Evolutions of the brane
universe described in both (26) and (29) can lead to the late-time
acceleration or deceleration depending on the parameters on the
model.

\begin{figure}[!hbtp] \label{fig2}
\begin{center}
\includegraphics[width=10cm]{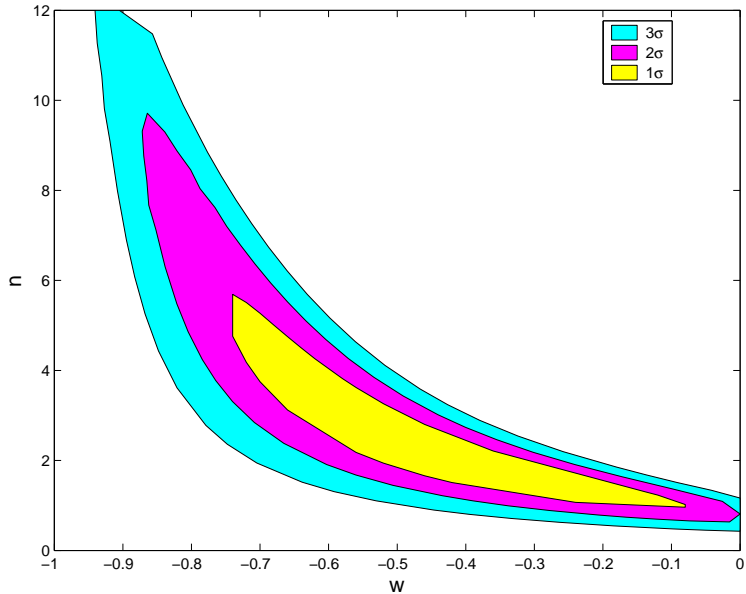}
\end{center}
\caption{The contour of the parameter space $w$-$n$.}
\end{figure}

We can define the deceleration parameter $q$ by using (\ref{eq26})
and (\ref{eq29}) and it is easy to see that $q$ can be negative in
both cases (see the solid line in Fig. 1). The effective dark
energy can be defined as
$\rho_{de}^{eff}=\rho_{de1}a^{-3(1+w)}+\rho_{de2}a^n$, where
$\rho_{de1}$ and $\rho_{de2}$ correspond to the coefficients of the
second and the third terms in (\ref{eq26}) and (\ref{eq29}),
respectively. The equation of state of the effective dark energy
can be defined by \cite{22}
\begin{equation}\label{eq30}
w_{de}^{eff}=-1-\frac{1}{3}\frac{d\ln \delta H^2}{d\ln a},
\end{equation}
where $\delta H^2=(H^2/H_0^2)-\Omega_m a^{-3}$. It is evolving
with time as shown in the dashed line in Fig1. Requiring the
deceleration parameter $q$ crossing $0$ around $0.5$ and the
effective equation of state of dark energy $w_{de}^{eff}$ crossing
$-1$ around $z=0.2$ as indicated by extensive analysis of
observational data \cite{17}, plus the flatness of the universe,
we can reduce the five parameters' space $(\rho_{m0}, C, w, n,
T_0)$ in (\ref{eq26}) and  $(D,\rho_{de0}, w, n, T_0)$ in
(\ref{eq29}) into two parameters' space $(w, n)$. Fitting the
newly released supernova legacy survey data~\cite{23}, the Sloan
Digital Sky Survey (SDSS) data~\cite{24} and the Wilkinson
Microwave Anisotropy Probe (WMAP) data~\cite{WMAP}, we obtain that
$w=-0.41^{+0.33}_{-0.34}$, $n=1.99^{+3.70}_{-1.07}$,
$\Omega_{m0}=0.28\pm 0.02$, $\Omega_{de1}=0.26^{+0.23}_{-0.15}$,
and $\Omega_{de2}=0.46^{+0.15}_{-0.24}$.  We see that in this
case, the dark energy with $w=-0.41$ on the brane will no longer
violate the causality condition. The contour of $w-n$ is shown in
Fig. 2. At the present the equation of state for the effective
dark energy is $w_{de0}^{eff}=-1.21$, which is obviously
consistent with current observation data~\cite{17,in3}.

In summary, we have investigated the role of the bulk-brane energy
exchange on the evolution of a brane universe. Due to the energy
flow between the bulk and the brane, the standard energy conservation
 is broken. Parameterizing the energy exchange term by
the scale factor and Hubble parameter, we have derived the
effective cosmological equations in the limit of low energy
density on the brane. Compared to the usual Friedmann equation, we
have got an additional modified term due to the brane-bulk energy
exchange. We found that the modified cosmological picture on the
brane accommodates the late time acceleration and the equation of
state for the effective dark energy experiences a transition
behavior from above $-1$ to below $-1$, while the dark energy on
the brane still obeys the causality condition. The ``phantom"
behavior of the effective dark energy is caused completely by
energy flow from the bulk. By fitting the recent type Ia
supernova, SDSS and WMAP data, we have shown that the modified
 gravity on the brane due to the bulk-brane energy exchange is
consistent with observations. Further, the effective dark energy equation of state
 $w_{de0}^{eff}=-1.21$ today shows that
our universe is under super-acceleration due to the considered
brane-bulk energy exchange. In addition, we note that due to the
energy exchange between the brane and the bulk, there will be a big
difference between the cases transferring the bulk energy to dark
matter and to dark energy on the brane in the effect on the  formation of large
scale structure. It is of great interest to further investigate
this issue.

\begin{acknowledgments}
This work was partially supported by  NNSF of China. Y. Gong
thanks the support by SRF for ROCS, State Education Ministry of
China. B. Wang's work was also supported by the Ministry of
Education of China and Shanghai Education Commission.
\end{acknowledgments}


\end{document}